\documentclass[letter]{aa}  

\usepackage{graphicx}
\usepackage{txfonts}
\usepackage{hyperref}
\usepackage{comment}
\hypersetup{colorlinks=true,allcolors=[rgb]{0,0,1}}
\usepackage[dvipsnames]{xcolor}

\usepackage{ulem}

\newcommand{\Teff}{$T_{\text{eff}}$}

\newcommand{\He}{$Y_{\text{He}}$}
\newcommand{\vsini}{$v \sin\!i$}

\newcommand{\kms}{${\rm km}\,{\rm s}^{-1}$}

\begin{document} 
\defcitealias{Mahy+22}{M+22}
\defcitealias{Barba+26}{B+26}
\defcitealias{Putkuri+26}{P+26}
\defcitealias{Trigueros-Paez+21}{TP+21}
\defcitealias{Frontera+01}{F+01}
\defcitealias{Chlebowski+89}{C+89}
\defcitealias{McSwain+11}{MS+11}
\defcitealias{Takahashi+09}{T+09}

   \title{The IACOB project}
   \subtitle{XVIII. Prevalence of short-period binaries among Galactic He-rich O stars}

   \author{C. Martínez-Sebastián\inst{1,2}
          \and
          R. Gamen
          \inst{3,4}
          \and
          O. G. Benvenuto
          \inst{3,4}
          \and
          E. A. Saavedra
          \inst{1,2}
          \and
          M. Carretero-Castrillo
          \inst{5}
          \and
          M. Contreras
          \inst{2}
          \and
          G. Holgado
          \inst{1,2}
          \and
          J. Maíz Apellániz
          \inst{6}
          \and
          S. Simón-Díaz
          \inst{1,2}}
          
   \institute{Instituto de Astrofísica de Canarias, c/Vía Láctea, S/N, E-38205 La Laguna, Tenerife, Spain
   \and
   Departamento de Astrofísica, Universidad de La Laguna, E-38206 La Laguna, Tenerife, Spain
   \and
   Instituto de Astrofísica de La Plata, CONICET–UNLP, Paseo del Bosque s/n, La Plata, Argentina
   \and 
   Facultad de Ciencias Astronómicas y Geofísicas, UNLP. Paseo del Bosque s/n, La Plata, Argentina
   \and 
   European Southern Observatory, Alonso de Córdova 3107, Vitacura, Santiago, Chile
   \and 
   Centro de Astrobiología. CSIC-INTA, Campus ESAC, C. bajo del castillo s/n, E-28 692 Villanueva de la Cañada, Madrid, Spain
   }
   \date{Received xx,xxxx; accepted xx,xxxx}

  \abstract
{For decades, the origin of helium enrichment in O-type stars has remained an open question.
In this study, we investigate the correlation between surface helium abundance (\He{}) and orbital parameters for a sample of 45 O-type SB1 systems --including Cyg X-1. We find seven He-rich systems, all of which are concentrated at short orbital periods ($P\!\lesssim\!15$~days) and are classified as runaways. In addition, four of them present relatively high eccentricities, while the other three have ellipsoidal variations. We argue that these properties are the result of binary interaction. These findings provide strong observational evidence that binary interaction is the dominant origin of helium enrichment in O-type stars. This result has important implications for the treatment of chemical mixing and surface abundances in massive-star evolutionary models, and establishes helium enrichment as a promising observational tracer for identifying post-interaction binary systems.}

   \keywords{stars: massive – 
             stars: abundances – 
             stars: evolution – 
             stars: atmospheres – 
             stars: binaries -
             binaries: spectroscopic
               }

   \maketitle

\vspace{-2cm}
\section{Introduction}\label{sec_intro}

Surface abundances provide key observational constraints to understand the evolution and internal physical processes of massive stars \citep[e.g.][]{Cantiello+07,Przybilla+08,Hunter+09,deMink+09,Kohler+12,song+13,McEvoy+15,Martins+17B,Martins+17A,Renzo&Gotberg21,Sen+22,Sciarini+26,Jin+26}. More than three decades ago, \cite{Herrero+92} showed that single-star evolution models available at that time could not reproduce the helium abundances observed in some OB stars, a problem they referred to as the ``helium discrepancy''. The inclusion of rotation in stellar evolution models initially appeared to resolve this issue \citep[e.g.][]{Meynet&Maeder00}. However, a new discrepancy emerged between nitrogen surface abundances and projected rotational velocities (\vsini{}), as the expected correlation between both quantities was not found for a remarkable number of stars \citep[e.g.][]{Hunter+08,Brott+11,Rivero-Gonzalez+12,Cazorla+17,Cazorla+17b}.

The accumulated challenges in the classical paradigm of single-star evolution led to a ``binary revolution'' \citep[i.e. the emphasis in the importance of binary interactions in massive stars; e.g.][]{Vanbeveren88}. \cite{Sana+12} marked a pivotal turning point showing that the majority of massive stars interact with a companion during their lifetime. This result introduced an additional complexity, as some observational properties --including those of apparently single stars-- may originate from past binary interactions \citep[][and references therein]{Langer12,Marchant&Bodensteiner24}. Since then, the community has devoted increasing effort to understanding the evolution of binary systems.

In this context, the MONOS \citep{Maiz-Apellaniz+19} and OWN \citep{Barba+17} surveys --covering the Northern and Southern hemispheres, respectively-- have focused on constraining the orbital properties of Galactic binaries with an O-type star. Complementarily, the IACOB project \citep[last described in][]{Simon-Diaz+20,Simon-Diaz2026} has exploited a large sample of high-resolution, high signal-to-noise spectra of Galactic massive stars to study their evolution in a holistic way. One of its key results is the evidence for a major role of binary interaction in the helium enrichment of Galactic O-type stars \citep{Martinez-Sebastian+25,Martinez-Sebastian+26,Simon-Diaz+26}, further supported by new model grids that follow the evolution of surface abundances of the accreting stars \citep{Jin+26}.

In this work, we combine orbital information with stellar parameters of the visible component for a sample of Galactic single-line spectroscopic binaries (SB1) containing an O-type star (Sect.~\ref{sec_sample}). We identify a clear correlation between helium enrichment, short orbital periods, and high eccentricities (Sect.~\ref{sec_results}). We interpret this behavior in the context of binary interaction, which emerges as the most plausible mechanism to explain the observed combination of properties (Sect.~\ref{sec_discussion}).

\vspace{-0.5cm}
\section{Sample and observational data}\label{sec_sample}

The sample analyzed in this work comprises 45 Galactic O-type stars identified as SB1 by the IACOB, MONOS, and/or OWN projects \citep{Simon-Diaz+11,Simon-Diaz+20,Maiz-Apellaniz+19,Trigueros-Paez+21,Barba+26}. For these objects, we compile (1) orbital parameters from \cite{Trigueros-Paez+21,Barba+26}, or \cite{Mahy+22} and (2) stellar parameters and surface helium abundances \citep{Holgado+25,Simon-Diaz+26}.
We list all targets in Table~\ref{tab_sample}, including their spectroscopic parameters --\vsini{}, effective temperature (\Teff{}), and surface helium abundance by number (\He{}\,=\,$N({\rm He})/N({\rm H})$)-- together with compiled orbital periods ($P$) and eccentricities ($e$). We also provide the spectroscopic binary status (SBS).
Additionally, we provide information on X-ray emission associated with these sources.

We present our SB1 classification criteria in App.~\ref{app_disentangling}. In short, we define SB2 systems as those in which a secondary component is clearly visible in at least one observed spectrum. When the secondary component is only revealed through spectral disentangling, we denote the system with an additional \textbf{d} flag in the SB classification. This approach differs from previous works that classify such cases as SB2 \citep[e.g.][]{Mahy+22}.

We classify He-rich stars following the criterion of \cite{Simon-Diaz+26}, i.e. stars with helium abundances exceeding the cosmic abundance standard from \cite{Nieva&Przybilla12} within uncertainties. This leads to \He{}\,>\,0.13. The remaining objects are classified as He-normal stars.
We complement the sample with information on the runaway (RW) status of each system.
We followed \cite{Carretero-Castrillo+23,Carretero-Castrillo+26} --i.e. using Gaia astrometric data to find objects that deviate significantly from the velocity distribution of normal stars--, while relaxing some of the selection criteria \citep[Table A.1 in][]{Carretero-Castrillo+23} to maximize the number of studied sources.
In Fig.~\ref{fig_sHR}, we show the distribution of the sample in the spectroscopic HR diagram \citep[sHRD;][]{Langer&Kudritzki14}.

\vspace{-0.5cm}
\section{Results}\label{sec_results}

\begin{figure}[!t]
\includegraphics[width=1.\hsize,trim={0 0 0 0}, clip]{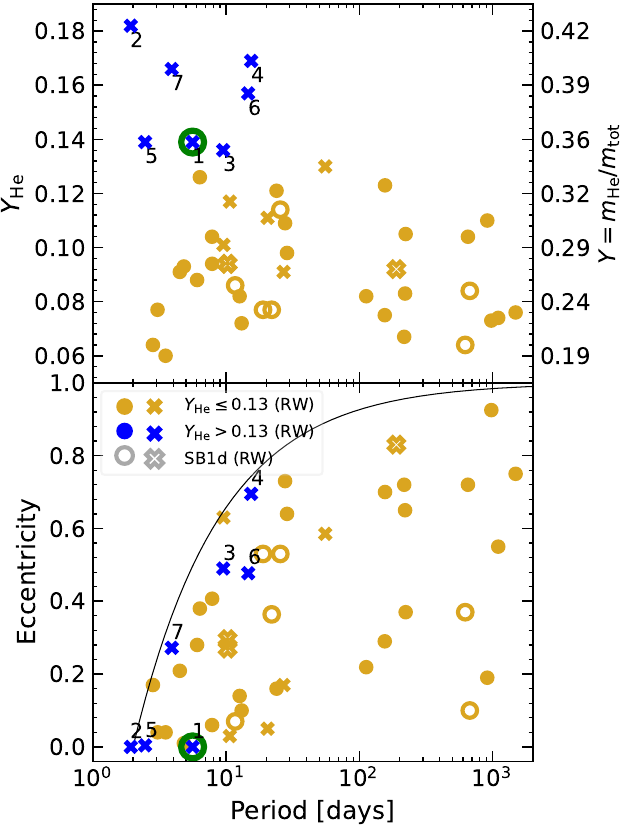}
  \caption{Surface He abundance and eccentricity as a function of the period. 
  Blue markers correspond to He-rich stars. He-normal stars are shown in yellow. Cross markers denote runaway stars. Open symbols indicate systems with a companion detected through disentangling (SB1d and d?). Numbers associated with He-rich stars indicate their order in Tables~\ref{tab_rw} and \ref{tab_sample}. 
  The green circle highlights HDE~226~868 (Cyg~X-1).}\label{fig_main}
  \vspace{-0.5cm}
\end{figure}

In the top panel of Fig.~\ref{fig_main}, we present the distribution of surface helium abundance as a function of the period.
First, we note that all He-rich systems have short periods ($P\!\lesssim\!15$~days), whereas He-normal stars show a broader distribution -- reaching $P\!\sim\!1000$~days, with only $\sim$45\% having $P\!\lesssim\!15$~days.
Moreover, all stars with \He{}\,$>\!0.13$ are identified as RW systems\footnote{We include Cyg~X-1 as a RW following the discussion in \cite{Carretero-Castrillo+23}.}. This contrasts with a RW incidence of $26\%$ among He-normal stars. 
We run a Fisher exact test \citep{Fisher56} to study the null hypothesis that the period and runaway distribution in both He-normal and He-rich samples are coming from the same parent distribution. With a $p$-value$<10^{-4}$, we rejected this possibility, showing a clear correlation between He-enrichment and these quantities.
Finally, none of the He-rich systems show evidence of a faint companion after disentangling, while $35\%$ of He-normal systems are SB1d.

In the bottom panel we present the orbital eccentricity as a function of orbital period. We find that He-rich systems tend to lie close to the period-eccentricity stability limit of \cite{Moe&diStefano17} (black solid line).

\vspace{-0.5cm}
\section{Discussion}\label{sec_discussion}

In this section, we argue that the observed properties of He-rich stars presented in this work are best explained by a binary-interaction origin, in line with \cite{Martinez-Sebastian+25,Martinez-Sebastian+26,Simon-Diaz+26}. In this scenario, the initially more massive star evolves more rapidly until it fills its Roche lobe. At this stage, it begins transferring mass to the companion. If sufficient mass is transferred, the accreting star gains material from progressively deeper, He-enriched layers, leading to a significant alteration of its surface composition (App.~\ref{app_models}).
In the following subsections, we discuss additional observed properties within the framework of binary interaction.

\subsection{The lack of long-period He-rich SB1 systems}\label{sec_longPeriod}

The proposed mass-transfer episode would significantly affect the orbital configuration, shrinking the orbit until the mass ratio is reversed. Simplified simulations indicate that systems with relatively low initial mass ratios ($q_0\!\sim\!0.6$) may reproduce the orbital periods observed in our sample (further discussion in App.~\ref{app_models}). However, these models also predict the presence of longer-period systems after mass-ratio reversal, which are not observed in our sample. We therefore investigated the possible origin of the lack of long-period He-rich SB1 systems. 

At long orbital periods, binary detection becomes increasingly challenging due to smaller radial-velocity (RV) variations. This limitation is particularly important for fast rotators (\vsini{}\,$\gtrsim\!200$~\kms{}), where line broadening reduces the precision of RV measurements \citep[e.g.][]{Britavskiy+23}. This effect may partly explain the lower detection rate of long-period SB1 systems with a fast rotator component (Fig.~\ref{fig_vsini} and App.~\ref{App_vsini}). However, the fact that $\sim\!50\%$ of He-normal systems are detected with periods longer than 20~days, while no He-rich systems are found in this regime, strongly suggests an intrinsic absence of long-period, slow-rotating He-rich binaries. If there are He-rich long period SB1, they should be fast rotators.
This is consistent with binary evolution after mass-ratio reversal, which leads to orbital widening and, consequently, a reduction in tidal efficiency. As a result, tidal forces may no longer prevent the spin-up of the accretor (see, however, discussion in App.~\ref{App_vsini}). Therefore, He-rich systems with $P\!\gtrsim\!15$~days may have originated in closer binaries and later evolved into wider, fast-rotating systems, making their spectroscopic detection more challenging. 

\cite{Simon-Diaz+26} reported a higher fraction of RW and a lower fraction of binaries among He-rich fast rotators (62\% and 12\%, respectively) compared to He-normal fast rotators (31\% and 21\%, respectively). Given the bias control in their study, this result is likely physical. In agreement with \cite{Britavskiy+23} and \cite{Carretero-Castrillo+26}, they interpret these runaways as products of the binary supernova scenario, in which a system is disrupted following the supernova (SN) explosion of the initially more massive star \citep{Blaauw1961}.

In summary, the absence of long-period, slow-rotating He-rich systems appears robust. While indirect evidence suggests a higher fraction of disrupted binaries among He-rich fast rotators \citep{Simon-Diaz+26}, this cannot yet be confirmed. Nevertheless, both scenarios that could explain the lack of long-period He-rich systems in our sample --either their evolution into fast rotators or their disruption through SN explosions-- are natural outcomes of binary evolution, further supporting the hypothesis that helium enrichment is driven by binary interaction.

\vspace{-0.3cm}
\subsection{The eccentricity-period relation of He-rich SB1 systems}

The evolution of the initially more massive star after mass transfer is expected to proceed relatively quickly. It evolves as a stripped star until it eventually undergoes a SN or collapses directly into a compact object (CO). In the former case, the event can significantly alter the orbital configuration -- even disrupting the system, as explained in the previous subsection.
Even when the SN does not disrupt the binary, it can still modify the orbit, increasing its eccentricity. In this context, a previous SN event would naturally explain why He-rich systems tend to occupy the upper region of the eccentricity–period stability relation shown in the bottom panel of Fig.~\ref{fig_main}: these would be systems that remain bound but whose orbits have been altered by the explosion of the currently unseen companion.

While tidal dissipation is expected to drive orbital re-circularization \citep{Zahn75}, the associated timescales can vary significantly\footnote{The effect of tides in eccentric systems is more complex than in circularized ones \citep[e.g.][]{Mardling&Aarseth01,Vick&Dong18}}. For systems such as Cyg~X-1, this process should be relatively fast ($\tau_{\rm circ}\!\sim\!9\!\cdot 10^3$~yr; \citealp{Schroder+21})\footnote{\cite{Mirabel&Rodrigues03} infer very low mass loss or a direct collapse in the SN that formed the black hole. In both cases, the post-supernova eccentricity would remain close to the pre-supernova value.}. However, for stars with radiative envelopes, circularization timescales can extend to a few million years \citep{Claret&Cunha97,Van_Eylen+16}.
Therefore, although mass transfer likely occurred in the past, some He-rich systems --particularly those with longer orbital periods-- may not yet have had sufficient time to re-circularize, and thus are still observed in eccentric orbits.

\vspace{-0.3cm}
\subsection{The runaway origin of He-rich SB1 systems}

Regarding the formation of RW, \cite{Renzo+19} found that the binary supernova scenario (BSS) is expected to produce lower two-dimensional peculiar velocities ($v^{\rm 2D}_{\rm PEC}$) with respect to the regional standard of rest than the dynamical ejection channel \citep[DES; ][]{Poveda+67}. Notably, BSS does not always lead to the disruption of the progenitor system \citep[e.g.][]{Fortin+22}.
\cite{Carretero-Castrillo+26} showed that different RW formation channels in Galactic O-type stars produce distinct distributions in \vsini{} and $v^{\rm 2D}_{\rm PEC}$ (Sect.~4.2 and Fig.~4 in the cited work). For our sample of He-rich systems, these quantities are summarized in Table~\ref{tab_rw}. Six out of seven systems $v^{\rm 2D}_{\rm PEC}\lesssim 50$~\kms{}, consistent with BSS products that remain bound after the SN \cite[e.g. Fig.~11 in][]{Renzo+19}.

For HDE~226~868 (Cyg X-1; star 1 in Fig.~\ref{fig_main}), a small kick is consistent with the results of \cite{Mirabel&Rodrigues03}. A similar interpretation may apply to HD~105~627 (star 3). Given their short orbital periods and eccentricities, a comparable scenario --potentially involving nearly direct core collapse-- may also explain HD~12~323 and HD~94~024 (stars 2 and 5). However, their comparatively large $v^{\rm 2D}_{\rm PEC}$ suggests either a highly asymmetric kick or, more plausibly, a contribution from the DES channel.
The combination of long orbital periods and moderate eccentricities in HD~14~633 and HD~130~298 (stars 4 and 6) suggests a larger mass-loss event, while their relatively low $v^{\rm 2D}_{\rm PEC}$ remains consistent with a BSS origin. In contrast, ALS~5039 (V479 Sct; star 7) is more likely a two-step product, reconciling its relatively large eccentricity for its orbital period --likely induced by the SN of the initially more massive star-- with its high $v^{\rm 2D}_{\rm PEC}$. 

We further note the absence of He-normal RW with $P\!<\!10$~days. 
This may indicate that short-period massive binaries that remain bound after the SN have typically undergone substantial mass transfer, leading to significant He-enrichment of the accretor.
Additional constraints on the evolutionary history of the He-rich systems may be obtained from the study of their companions, which are expected to be CO or stripped stars (a brief discussion is presented in App.~\ref{App_companion}; a detailed analysis is beyond the scope of this work), as well as from other observables such as the projected rotational velocity (App.~\ref{App_vsini}).

\begin{table}[]
\tiny
    \centering
    \caption{Projected rotational velocity and 2-D peculiar velocities \citep[$v^{\rm 2D}_{\rm PEC}$; ][]{Carretero-Castrillo+26} for the He-rich systems.}
    \label{tab_rw}
    \begin{tabular}{lll}
         \hline\hline
         Star & \vsini{} & $v^{\rm 2D}_{\rm PEC}$ \\
         & [\kms{}] & [\kms{}] \\
         \hline
        HDE~226~868 (Cyg~X-1) & 95 & 21.6\\
        HD~12~323 & 121 & 53.4\\
        HD~105~627 & 141 & 16.4\\
        HD~14~633 & 121& 31.1\\
        HD~94~024 & 162 & 49.9\\
        HD~130~298 & 167 & 30.4\\
        ALS~5039 (V479~Sct) & 124 & 90.9\\
        \hline
    \end{tabular}
    \tablefoot{Same order as Table~\ref{tab_sample}.}
\end{table}

\vspace{-0.5cm}
\section{Conclusions}\label{sec_conclusions}

In this work, we have analyzed the properties of 45 SB1 systems, combining surface helium abundances with their orbital solutions. We find that all He-rich systems in our sample are concentrated at short orbital periods ($P\!\lesssim\!15$~days) and lie close to the maximum eccentricity compatible with bound systems. In addition, all of them are classified as runaways, with relatively low peculiar velocities (six out of seven with $v^{\rm 2D}_{\rm PEC} \lesssim\!50$~\kms{}). Finally, we do not detect signatures of companions in any of these systems after spectral disentangling.

We propose that these characteristics are the result of binary interaction. In this scenario, the initially more massive star transfers mass to its companion, which we now observe as the primary, while the donor has evolved into a stripped star or CO. Part of the accreted material originates from the inner layers of the donor, enriched in processed elements, including helium, thereby modifying the surface composition of the accretor. The observed short orbital periods are consistent with initially low mass ratios.

The subsequent evolution of the donor may lead to a SN explosion, which can both impart a systemic velocity (producing a RW in the BSS) and modify the orbital configuration by increasing its eccentricity or even disrupting the system. The former scenario is consistent with the observed incidence of RW, the combination of \vsini{} and $v^{\rm 2D}_{\rm PEC}$ among He-rich systems, and their high eccentricities. The latter is compatible with the lower binary fraction and higher RW incidence among He-rich stars reported by \cite{Simon-Diaz+26} \citep[see also Sect.~4.2 in ][]{Blaauw93}.

This study adds to the growing body of observational evidence from the IACOB project indicating that helium enrichment in Galactic O-type stars is most likely the result of prior binary interaction \citep{Martinez-Sebastian+25,Martinez-Sebastian+26,Simon-Diaz+26}. These results have important implications for both single- and binary-star evolutionary models, providing constraints on key parameters such as mass-transfer efficiency. In this context, He-rich O-type stars emerge as valuable laboratories for studying binary interaction and its subsequent evolution.

Nonetheless, several questions remain open. In particular, the apparent absence of He-rich fast rotators requires further investigation within the binary-interaction framework. Additionally, constraining the nature of the unseen companions in He-rich systems -- potentially through multi-wavelength observations; see App.~\ref{App_Xray} -- and studying their dynamical environments will be crucial to further advance our understanding of the evolution of massive binary (and multiple) systems.


\bibliographystyle{aa}
\bibliography{references}

\begin{appendix}

\section{Sample parameters and distribution}

\begin{figure}[!t]
\includegraphics[width=\hsize,trim={0 1 0 0}]{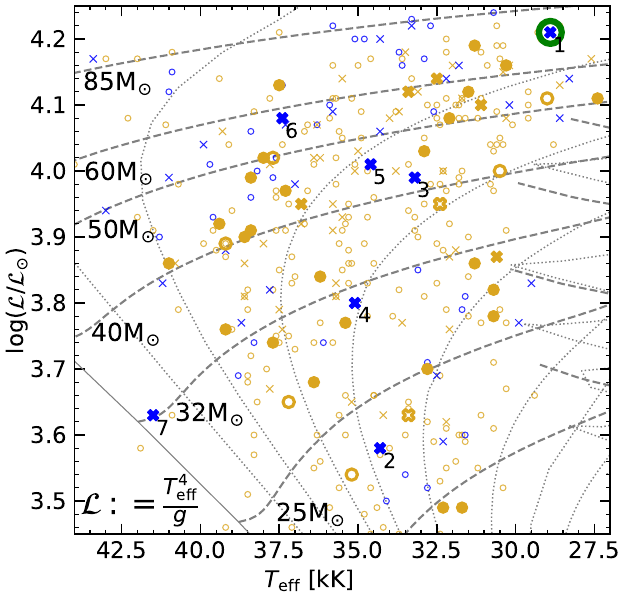}
  \caption{Location of the 45 SB1 systems in the sHRD (large symbols). 
  Same coding as in Fig.~\ref{fig_main}.
  As references, we include GENEC evolutionary tracks for the main sequence without initial rotation (dashed lines; \citealt{Ekstrom+12}) and SYCLIST isochrones in steps of 1~Myr (dotted lines; \citealt{Georgy+14}). The full sample from \cite{Simon-Diaz+26} is shown in the background as small open circles.}
\label{fig_sHR}
\end{figure}

In Table~\ref{tab_sample} we provide the spectral classification \citep[from GOSSS; ][]{Maiz-Apellaniz+11}, spectral parameters from the visible star \cite[projected rotational velocity, effective temperature, gravity and spectroscopic luminosity\footnote{$\mathcal{L}\!:=\!T_{\rm eff}^4/g$; \cite{Langer&Kudritzki14}}; ][]{Simon-Diaz+26}, orbital period and eccentricity, X-ray flux (if available), and SB status of the studied sample (App.~\ref{app_disentangling}).
First seven rows correspond to He-rich systems. 
In Fig.~\ref{fig_sHR} we present the distribution of the SB1 sample (large symbols) over the full sample of O-type stars from the IACOB project (small symbols) in the spectroscopic HR diagram. He-rich systems (blue markers) are numbered according to their corresponding row in Table~\ref{tab_sample}.

\begin{table*}
\tiny
    \centering
    \caption{Information of the studied sample.}
    \label{tab_sample}
    \begin{tabular}{llrlllrllllll}
         \hline\hline \\
         Star & SpC & 
\vsini{} & \Teff{} & $\log \mathcal{\frac{L}{L_{\odot}}}$ & \He{}$\times100$ & $P$ & $e$ & Ref. & X-ray flux & Ref. & SBS \\
 & & [\kms{}] & [kK] & [dex] & & [days] & & orb. & [${\rm erg}\,{\rm s}^{-1}\,{\rm cm}^{-2}$] & X-ray &\\
\hline
HDE~226~868 &  O9.7~Iabp~var  & 95 & 28.9\,$\pm$\,0.6 & 4.2\,$\pm$\,0.1 & 13.9\,$\pm$\,5.5 &  5.6 &  0.0  &  \citetalias{Trigueros-Paez+21}  & $(7.1\pm2.9)\!\cdot\!10^{-8}$ & \citetalias{Frontera+01} &  SB1E$^{\dagger}$ \\
HD~12~323 &  ON9.2~V  & 121 & 34.3\,$\pm$\,0.9 & 3.6\,$\pm$\,0.2 & 18.2\,$\pm$\,5.1 &  1.9  &  0.0  &  \citetalias{Trigueros-Paez+21}  & $\sim5\!\cdot\!10^{-13}$ & \citetalias{Chlebowski+89} &  SB1E$^{\dagger}$ \\
HD~105~627 &  O9~III  & 141 & 33.2\,$\pm$\,0.6 & 4.0\,$\pm$\,0.1 & 13.6\,$\pm$\,3.9 &  9.5  &  0.49  &  OWN   & $2.4^{+1.1}_{-0.9}\!\cdot\!10^{-14}$ & eR &  SB1 \\
HD~14~633 &  ON8.5~V  & 121 & 35.1\,$\pm$\,0.5 & 3.8\,$\pm$\,0.1 & 16.9\,$\pm$\,3.5 &  15.4  &  0.695  &  \citetalias{Trigueros-Paez+21}  & $1.1_{-0.7}^{+1.7}\!\cdot\!10^{-13}$ & \citetalias{McSwain+11} &  SB1 \\
HD~94~024 &  O8~IV  & 162 & 34.6\,$\pm$\,0.6 & 4.0\,$\pm$\,0.0 & 13.9\,$\pm$\,3.9 &  2.5  &  0.004  &  \citetalias{Barba+26}   & $2.9^{+1.5}_{-1.1}\!\cdot\!10^{-14}$ & eR &  SB1E$^{\dagger}$ \\
HD~130~298 &  O6.5~III(n)(f)  & 167 & 37.4\,$\pm$\,0.8 & 4.1\,$\pm$\,0.1 & 15.7\,$\pm$\,2.6 &  14.6  &  0.477  &  \citetalias{Barba+26}   & $<1.4\!\cdot\!10^{-13}$ & eR &  SB1 \\
ALS~5039 &  ON6~V((f))z  & 124 & 41.5\,$\pm$\,1.5 & 3.6\,$\pm$\,0.1 & 16.6\,$\pm$\,0.1 &  3.9  &  0.272  &  \citetalias{Trigueros-Paez+21}  & $(8.3\pm3.4)\!\cdot\!10^{-12}$ & \citetalias{Takahashi+09} &  SB1 \\
\hline
HDE~326~329 &  O9.7~V  & 87 & 32.3\,$\pm$\,0.9 & 3.5\,$\pm$\,0.1 & 12.3\,$\pm$\,2.9 &  155.7  &  0.7  &  OWN & $<4.1\!\cdot\!10^{-13}$ & eR &  SB1 \\
HD~308~813 &  O9.7~IV(n)  & 215 & 31.7\,$\pm$\,0.7 & 3.5\,$\pm$\,0.1 & 12.6\,$\pm$\,2.6 &  6.4  &  0.38  &  \citetalias{Mahy+22}   & $<7.2\!\cdot\!10^{-14}$ & eR &  SB1 \\
HD~154~643 &  O9.7~III  & 101 & 31.3\,$\pm$\,0.8 & 3.9\,$\pm$\,0.1 & 9.8\,$\pm$\,3.8 &  28.6  &  0.64  &  \citetalias{Barba+26}  & $8.9^{+3.2}_{-2.7}\!\cdot\!10^{-14}$ & eR &  SB1 \\
HD~165~174 &  O9.7~IIn  & 264 & 30.7\,$\pm$\,1.2 & 3.8\,$\pm$\,0.1 & 12.1\,$\pm$\,4.5 &  23.9  &  0.16  &  \citetalias{Mahy+22}   & -- &--&  SB1 \\
HD~152~405 &  O9.7~II  & 59 & 30.5\,$\pm$\,0.7 & 4.0\,$\pm$\,0.1 & 11.4\,$\pm$\,4.0 &  25.5  &  0.55  &  \citetalias{Mahy+22}   & $5.0^{+2.5}_{-1.2}\!\cdot\!10^{-14}$ & eR &  SB1d? \\
HD~167~264 &  O9.7~Iab  & 71 & 29.0\,$\pm$\,0.5 & 4.1\,$\pm$\,0.1 & 8.4\,$\pm$\,2.2 &  673.4  &  0.1  &  OWN-II   & -- &--&  SB1d \\
BD~+36~4063 &  ON9.7~Ib  & 118 & 27.4\,$\pm$\,1.0 & 4.1\,$\pm$\,0.1 & 9.3\,$\pm$\,3.5 &  4.8  &  0.01  &  \citetalias{Trigueros-Paez+21}  & -- &--&  SB1E$^{\dagger}$ \\
HD~192~001 &  O9.5~IV  & 44 & 33.4\,$\pm$\,0.9 & 3.6\,$\pm$\,0.1 & 9.2\,$\pm$\,2.2 &  189.4  &  0.83  &  \citetalias{Mahy+22}   & -- &--&  SB1d? \\
HD~163~892 &  O9.5~IV(n)  & 215 & 32.8\,$\pm$\,0.6 & 3.7\,$\pm$\,0.1 & 10.4\,$\pm$\,2.1 &  7.8  &  0.06  &  \citetalias{Barba+26}   & -- &--&  SB1 \\
HD~37~737 &  O9.5~II-III(n)  & 201 & 30.7\,$\pm$\,0.8 & 3.8\,$\pm$\,0.1 & 9.4\,$\pm$\,3.3 &  7.9  &  0.407  &  \citetalias{Trigueros-Paez+21}  & -- &--&  SB1E \\
HD~15~137 &  O9.5~II-IIIn  & 270 & 30.6\,$\pm$\,0.9 & 3.9\,$\pm$\,0.1 & 13.0\,$\pm$\,5.1 &  55.4  &  0.5852  &  \citetalias{Trigueros-Paez+21}  & $(1-2)\!\cdot\!10^{-14}$ & \citetalias{McSwain+11} &  SB1 \\
HD~164~438 &  O9.2~IV  & 55 & 32.4\,$\pm$\,0.8 & 4.0\,$\pm$\,0.0 & 9.4\,$\pm$\,2.8 &  10.3  &  0.296  &  \citetalias{Trigueros-Paez+21}  & -- &--&  SB1d? \\
HD~76~968 &  O9.2~Ib  & 53 & 31.1\,$\pm$\,0.7 & 4.1\,$\pm$\,0.1 & 9.1\,$\pm$\,3.0 &  27.0  &  0.17  &  \citetalias{Barba+26}  & $1.2^{+0.4}_{-0.3}\!\cdot\!10^{-13}$ & eR &  SB1 \\
HD~152~424 &  OC9.2~Ia  & 59 & 30.3\,$\pm$\,0.8 & 4.2\,$\pm$\,0.1 & 6.7\,$\pm$\,1.7 &  216.8  &  0.72  &  \citetalias{Barba+26} & $1.8\pm0.4\!\cdot\!10^{-13}$ & eR &  SB1 \\
HDE~229~234 &  O9~III  & 97 & 32.1\,$\pm$\,0.9 & 4.1\,$\pm$\,0.1 & 6.0\,$\pm$\,0.0 &  3.5  &  0.04  &  \citetalias{Trigueros-Paez+21}  & -- &--&  SB1E$^{\dagger}$ \\
HD~57~061 &  O9~II  & 57 & 32.9\,$\pm$\,1.1 & 4.0\,$\pm$\,0.1 & 7.5\,$\pm$\,2.2 &  155.0  &  0.29  &  \citetalias{Barba+26}  & $1.8\pm0.1\!\cdot\!10^{-12}$ & eR &  SB1 \\
HD~16~429 &  O9~II(n)  & 193 & 31.5\,$\pm$\,0.8 & 4.1\,$\pm$\,0.1 & 7.7\,$\pm$\,2.4 &  3.1  &  0.04  &  \citetalias{Trigueros-Paez+21}  & -- &--&  SB1 \\
HD~73~882 &  O8.5~IV  & 145 & 35.4\,$\pm$\,0.8 & 3.8\,$\pm$\,0.1 & 7.2\,$\pm$\,1.3 &  13.1  &  0.1  &  \citetalias{Barba+26}  & $1.8^{+0.4}_{-0.3}\!\cdot\!10^{-13}$ & eR &  SB1 \\
HD~52~533 &  O8.5~IVn  & 299 & 35.2\,$\pm$\,0.5 & 3.5\,$\pm$\,0.1 & 7.7\,$\pm$\,1.1 &  22.0  &  0.364  &  \citetalias{Trigueros-Paez+21}  & $1.4\pm0.4\!\cdot\!10^{-13}$ & eR &  SB1Ed \\
HD~75~211 &  O8.5~II((f))  & 145 & 33.4\,$\pm$\,0.7 & 4.1\,$\pm$\,0.1 & 11.1\,$\pm$\,2.2 &   20.5  &  0.05  &  \citetalias{Barba+26}   & $1.6\pm0.4\!\cdot\!10^{-13}$ & eR &  SB1 \\
HD~74~194 &  O8.5~Ib-II(f)p  & 184 & 32.5\,$\pm$\,0.7 & 4.1\,$\pm$\,0.1 & 10.1\,$\pm$\,1.6 &  9.5  &  0.63  &  \citetalias{Barba+26}   & $2.4^{+0.6}_{-0.3}\!\cdot\!10^{-13}$ & eR &  SB1 \\
HD~112~244 &  O8.5~Iab(f)p  & 124 & 31.3\,$\pm$\,0.8 & 4.2\,$\pm$\,0.1 & 10.9\,$\pm$\,3.2 &  27.7  &  0.73  &  \citetalias{Barba+26}  & $7.1\pm0.6\!\cdot\!10^{-13}$ & eR &  SB1 \\
HD~101~413 &  O8~V  & 80 & 36.4\,$\pm$\,0.8 & 3.7\,$\pm$\,0.1 & 7.3\,$\pm$\,1.5 &  977.2  &  0.93  &  \citetalias{Putkuri+26} & $6.1^{+3.4}_{-1.4}\!\cdot\!10^{-14}$ & eR &  SB1d \\
HD~152~590 &  O7.5~Vz  & 48 & 37.7\,$\pm$\,0.7 & 3.7\,$\pm$\,0.1 & 9.1\,$\pm$\,1.7 &  4.5  &  0.209  &  OWN-II & $6.4^{+2.7}_{-2.2}\!\cdot\!10^{-14}$ & eR &  SB1Ed \\
HD~164~536 &  O7.5~V(n)  & 237 & 37.2\,$\pm$\,0.7 & 3.6\,$\pm$\,0.1 & 8.6\,$\pm$\,1.7 &  11.7  &  0.07  &  \citetalias{Mahy+22}   & -- &--&  SB1d? \\ 
HD~46~573 &  O7.5~V((f))  & 77 & 36.8\,$\pm$\,0.7 & 4.0\,$\pm$\,0.1 & 11.7\,$\pm$\,2.2 &  10.7  &  0.03  &  \citetalias{Trigueros-Paez+21}  & $8.13^{+4.02}_{-2.93}\!\cdot\!10^{-14}$ & eR &  SB1 \\
HD~91~824 &  O7~V((f))z  & 51 & 39.2\,$\pm$\,0.7 & 3.8\,$\pm$\,0.1 & 8.2\,$\pm$\,1.6 &  112.5  &  0.22  &  OWN-II   & $1.20^{+0.25}_{-0.22}\!\cdot\!10^{-13}$ & eR &  SB1d \\
HD~93~146 &  O7~V((f))  & 60 & 38.4\,$\pm$\,0.7 & 4.0\,$\pm$\,0.1 & 7.4\,$\pm$\,1.4 &  1101.0  &  0.55  &  OWN-II & $<2.1\!\cdot\!10^{-13}$ & eR &  SB1d \\
HD~101~205 &  O7~II:(n)  & 316 & 36.2\,$\pm$\,0.8 & 3.8\,$\pm$\,0.1 & 6.4\,$\pm$\,1.0 &  2.8  &  0.17  &  \citetalias{Barba+26}  & $6.2^{+0.5}_{-0.4}\!\cdot\!10^{-13}$ & eR &  SB1E \\
HD~91~572 &  O6.5~V((f))z  & 60 & 38.4\,$\pm$\,0.7 & 3.9\,$\pm$\,0.1 & 7.6\,$\pm$\,1.7 &  1487.0  &  0.75  &  \citetalias{Barba+26}  & $1.4^{+0.3}_{-0.2}\!\cdot\!10^{-13}$ & eR &  SB1 \\
HD~322~417 &  O6.5~IV((f))  & 68 & 38.0\,$\pm$\,1.0 & 4.0\,$\pm$\,0.1 & 10.5\,$\pm$\,3.2 &  222.3  &  0.37  &  \citetalias{Barba+26}  & $1.3^{+0.4}_{-0.3}\!\cdot\!10^{-13}$ & eR &  SB1 \\
HD~96~946 &  O6.5~III(f)  & 72 & 38.6\,$\pm$\,0.9 & 3.9\,$\pm$\,0.1 & 11.0\,$\pm$\,3.0 &  910.9  &  0.19  &  \citetalias{Barba+26}  & $<6.2\!\cdot\!10^{-14}$ & eR &  SB1 \\
HD~152~723 &  O6.5~III(f)  & 73 & 37.7\,$\pm$\,0.7 & 4.0\,$\pm$\,0.1 & 7.7\,$\pm$\,1.2 &  18.9  &  0.51  &  \citetalias{Mahy+22}   & $3.7^{+0.6}_{-0.5}\!\cdot\!10^{-13}$ & eR &  SB1d? \\
CPD~-59~2600 &  O6~V((f))  & 127 & 39.2\,$\pm$\,1.1 & 3.9\,$\pm$\,0.1 & 6.4\,$\pm$\,1.0 &  622.6  &  0.37  &  \citetalias{Barba+26}  & $1.7\pm0.3\!\cdot\!10^{-13}$ & eR &  SB1d \\
HD~101~190 &  O6~IV((f))  & 49 & 39.4\,$\pm$\,0.9 & 3.9\,$\pm$\,0.1 & 8.8\,$\pm$\,1.6 &  6.1  &  0.28  &  \citetalias{Barba+26}  & $2.8\pm0.3\!\cdot\!10^{-13}$ & eR &  SB1 \\
HD~124~314 &  O6~IV(n)((f))  & 256 & 37.3\,$\pm$\,0.8 & 4.0\,$\pm$\,0.1 & 8.3\,$\pm$\,1.3 &  220.1  &  0.65  &  \citetalias{Barba+26}  & $4.7\pm0.6\!\cdot\!10^{-13}$ & eR &  SB1 \\
HDE~319~699 &  O5~V((fc))  & 69 & 41.0\,$\pm$\,1.3 & 3.9\,$\pm$\,0.1 & 8.2\,$\pm$\,2.0 &  12.6  &  0.14  &  \citetalias{Barba+26}  & $2.5\pm0.5\!\cdot\!10^{-13}$ & eR &  SB1 \\
CPD~-47~2963 &  O5~Ifc  & 67 & 37.5\,$\pm$\,0.9 & 4.1\,$\pm$\,0.1 & 10.4\,$\pm$\,2.4&  652.5  &  0.72  &  \citetalias{Barba+26}  & $2.9\pm0.5\!\cdot\!10^{-13}$ & eR &  SB1 \\
\hline
    \end{tabular}
    \tablefoot{Columns include: star identifier; spectral classification (SpC) from the ALS catalog; projected rotational velocity (\vsini{}), effective temperature (\Teff{}), spectroscopic luminosity $\log \mathcal{L}/\mathcal{L}_{\odot}$ \citep{Langer&Kudritzki14}, and surface He abundance in number (\He{}) from the IACOB project \citep{Simon-Diaz+26}; orbital period and eccentricity, together with their corresponding references; X-ray flux and its reference (``eR'' when obtained from eROSITA); and binary status (SB; see text for details).
    We denote systems with ellipsoidal variations with ``$^{\dagger}$'' and systems with a companion resulting from disentangling with ``d'' after after the SB classification.}\\
    \tablebib{(\citetalias{Trigueros-Paez+21}) \citet{Trigueros-Paez+21}; 
    (\citetalias{Mahy+22}) \citet{Mahy+22};  (\citetalias{Barba+26}) \citet{Barba+26}; (\citetalias{Putkuri+26}) \citet{Putkuri+26};   
    (\citetalias{Chlebowski+89}) \citet{Chlebowski+89}; \citetalias{Frontera+01}) \citet{Frontera+01}; (\citetalias{Takahashi+09}) \citet{Takahashi+09}; (\citetalias{McSwain+11}) \citet{McSwain+11}}
\end{table*}

\section{Nature of the unseen companions}

\subsection{SB classification and spectra disentangling} \label{app_disentangling}

Single-lined binaries (SB1) are classically defined as ``a spectroscopic binary in which only one set of spectral lines is detectable'' \citep{Ridpath04}. However, the recent development and extensive use of different disentangling techniques have raised a debate: through disentangling, it is now possible to reveal signatures of previously hidden companions --either because they are significantly fainter than the visible component or because they are fast rotators with strongly broadened spectral lines. 

The application of this technique to modern multi-epoch spectroscopic datasets has opened new avenues for the study of binary systems, enabling a more comprehensive characterization of their components. Among other advantages, it allows the independent analysis of the spectra and orbital motion of each star in the system.
As a consequence, some recent studies classify systems as SB2 when signatures of the secondary component are recovered through spectral disentangling \citep[e.g.][]{Mahy+22}.

However, this approach raises major concerns. 
SB1 systems were historically defined based on the direct analysis of the observed spectra --without the use of disentangling techniques. Modifying this definition could complicate the comparison of results from the existing literature. 
Moreover, the advanced analysis required for spectral disentangling may lead to inhomogeneous classifications in large samples, where only a subset of the targets undergo disentangling. As a consequence, different criteria may be implicitly adopted within the same work to classify systems as SB1 or SB2, biasing possible interpretations.
Finally, the results of this technique are sensitive to several factors, including the adopted approach, the initial guess, and the orbital solution, which may lead to discrepant detections. HD~163~892 provides an illustrative example of this issue: while \cite{Mahy+22} reported a companion, the reanalysis by \cite{Barba+26} found no evidence of it.

With these concerns in mind, we opted to classify systems as SB2 only when the secondary component is qualitatively detectable in at least one of the available spectra without the need for additional analysis. In particular, our classification is based on the presence of a secondary signature in He~{\sc I}~5875 and/or the O~{\sc I}~7772--7775 triplet.
Consequently, even if a companion is recovered through disentangling, we still classify the system as SB1 when no secondary component is directly visible. This criterion differs from that adopted by \cite{Mahy+22}, among others, but follows the spirit of the approach advocated by N. Walborn, who argued that classification must not rely on measurements.

Nonetheless, we recognize the importance of detecting faint companions. For this reason, we implemented an additional qualifier to the classification introduced in \cite{Maiz-Apellaniz+19}. Namely, we add a \textbf{d} to the SB status whenever a companion is detected after disentangling.
In cases where the analysis either (i) reveals only very weak signatures of a companion --barely above the noise level-- or (ii) does not reveal any companion despite previous claims in the literature, we assign the qualifier \textbf{d?}.

To homogenize our sample, we performed spectral disentangling with UNWIND \citep{Maiz-Apellaniz+26} for all systems. Although this analysis had already been carried out by \cite{Barba+26}, we revisited the systems from \cite{Trigueros-Paez+21} and \cite{Mahy+22}, as the latter were analyzed using a different tool.
We also reanalyzed the available TESS photometry for the sample. To further constrain the possible nature of the companions, we included information on eclipses (\textbf{E}) and ellipsoidal variability (\textbf{E$^{\dagger}$}) in the SB classification. The results of these analyses are summarized in the last column of Table~\ref{tab_sample}.
In the figures of this work, we have consider SB1d? systems as SB1d (empty markers).

\subsection{X-ray emission}\label{App_Xray}

To gather additional information about the unseen companions in our SB1 systems, we searched for X-ray counterparts associated with each source.
Several mechanisms can produce X-ray emission in O-type binaries: intrinsic wind-embedded shocks generated by the line-deshadowing instability \citep[][]{Lucy+80,Owocki+88,Feldmeier+97,Naze+09}; colliding stellar winds in systems containing two non-degenerate stars with significant winds \citep{Stevens+92,Rauw+16}; magnetically confined wind shocks in magnetic massive stars \citep{Babel+92,ud-Doula+16}; and accretion onto a compact companion --white dwarf, neutron star, or black hole-- fed by the primary stellar wind \citep{Davidson+73,Martinez+17}.
In our sample, the colliding-wind scenario is disfavored for systems lacking spectroscopic signatures of a luminous secondary (i.e. systems not classified as SB1d), while magnetic confinement is unlikely given the absence of confirmed magnetic detections among our targets \citep{Grunhut+17,Fossati+15}. Therefore, X-ray detections above the level expected from intrinsic wind emission may indicate the presence of a compact companion.

We compiled X-ray fluxes --or upper limits-- from both the literature and the first data release (DR1) of the \textit{eROSITA} all-sky survey \citep{Merloni+24}. For DR1 detections, we adopted the catalogue flux in the 0.2--2.3 keV band. For sources not detected in DR1, we adopted the corresponding upper limit in the same band. These upper limits are model-dependent, as their count-rate-to-flux conversion relies on an assumed absorbed power-law spectral model \citep{TubinArenas+24}. Literature fluxes were taken from the original references in their quoted energy bands and flux definitions. The results are presented in the ``X-ray flux'' column of Table~\ref{tab_sample} (-- indicates that no data are available). This gives extra information on the possible companions of the objects (App.~\ref{App_companion}).

\subsection{Possible companions of He-rich systems}\label{App_companion}

The He-rich systems in our sample exhibit relative high eccentricities, runaway characteristics, and a \vsini{} distribution broadly consistent with synchronous rotation. These properties make them strong candidates for having experienced a supernova event involving the initially more massive component (Sect.~\ref{sec_discussion}). If this scenario is correct, these systems would be promising candidates to host CO companions.
Interestingly, all He-rich systems show X-ray detections (Table~\ref{tab_sample}), which is not expected for binaries hosting non-degenerate companions (App.~\ref{App_Xray}).
Alternatively, systems with the shortest orbital periods and low eccentricities may represent post-interaction, pre-supernova configurations hosting a stripped-star companion \citep[e.g.][]{Drout+23,Gotberg+23,Naze&Gregor25}. In this scenario, the RW nature of the binary could result from the dynamical ejection of the system \citep[e.g.][]{Oh&Kroupa16}.
However, this scenario appears less likely, as the post-mass-transfer evolution of the donor is expected to proceed relatively rapidly (App.~\ref{app_models}).

As the He-rich systems analyzed in this work are also part of the sample studied by \cite{Mahy+22}, we have complementary constraints on the nature of their unseen companions. In particular, they estimated the companion masses as a function of the orbital inclination and the possible mass range for different companions (their Figs.~6 and 9, respectively).

At present, only HDE~226~868 (Cyg~X-1) is confirmed to harbor a black-hole companion \citep{Miller-Jones+21}. Nonetheless, HD~12~323 and HD~94~024 are also candidates. Both systems display periodic variations in their TESS light curves at half their orbital periods, consistent with ellipsoidal variability. \cite{Britavskiy+23} proposed both of them as candidate black-hole binaries. However, \cite{Mahy+22} derived companion masses more consistent with neutron stars($M_{\rm 2, min}\!=\!1.3\!\pm\!0.3~M_{\odot}$ and $1.4\!\pm\!0.2~M_{\odot}$, respectively). Otherwise, these masses could also be compatible with stripped-star companions. In the latter case, the companions could produce an excess of ultraviolet emission \citep{Gotberg+18}. Observations in the UV domain would therefore be valuable to test this possibility.

\cite{Mahy+22} proposed HD~130~298 as the most promising black-hole candidate in their sample --based on its RV semi-amplitude and the minimum detectable mass of the companion; see also \cite{Sana+22}.
Together with HD~14~633, these systems exhibit long orbital period and high eccentricity, suggesting significant mass loss during their evolution. Their relatively low $v^{\rm 2D}_{\rm PEC}$ values are consistent with peculiar velocities produced through the BSS, leading to the formation of a black hole in the former system \citep{Mahy+22} and potentially a neutron star in the latter \citep{McSwain+07,McSwain+11}.
In contrast, HD~105~627 shows a very low $v^{\rm 2D}_{\rm PEC}$ together with relatively short orbital period and low eccentricity, consistent with a nearly direct core-collapse event. Its companion has also been proposed to be a compact object \citep[][and references therein]{Dubas13,Mahy+22}.

Finally, ALS~5039 (V479 Sct) is a well-known gamma-ray binary hosting a compact-object companion \citep{Casares+05}, most likely a neutron star \citep{Yoneda+20,Volkov+21}. The system combines a short orbital period with a relatively high eccentricity.
A two-step ejection scenario could reconcile its large eccentricity --produced during the supernova explosion of the initially more massive star-- with its high $v^{\rm 2D}_{\rm PEC}$. This interpretation would be consistent with substantial mass loss during the supernova event \citep{Ribo+02}.

\section{Interacting binary models}\label{app_models}

To interpret the observations presented in the previous sections, we analyzed the models presented by \citet{Jin+26}\footnote{The stellgrid web service is available at \url{https://wwwmpa.mpa-garching.mpg.de/stellgrid/}.}. In this work, we analyze only a model to demonstrate the plausibility and robustness of our proposed scenario. A more detailed analysis of the observations within the framework of binary evolution models will be the subject of future work.

\begin{figure}
\centering
\includegraphics[width=1\columnwidth]{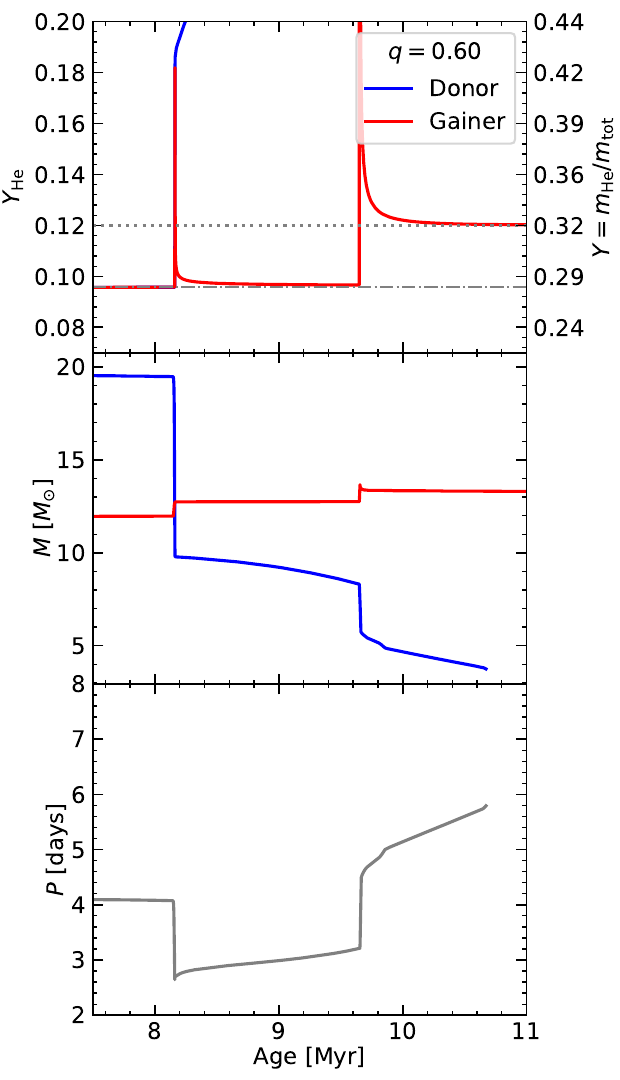}
    \caption{Upper panel: evolution of the surface He abundance of the two components of a binary system initially composed of solar-composition stars with masses of $M_{1}\!=\!19.95~M_{\odot}$ (donor) and $M_{2}\!=\!11.97~M_{\odot}$ (gainer), in a circular orbit with an initial period of $P\!=\!3.98$~days and an initial rotational velocity of $v_{\rm rot}\!=\!100$~\kms{}. Middle panel: evolution of the component masses. Lower panel: evolution of the orbital period. The dashed horizontal line in the upper panel indicates the final surface helium abundance of the gainer. The resulting helium enrichment and orbital period are compatible with the observations.}
    \label{fig_modelsMPA}
\end{figure}

We selected a solar-composition model consisting initially of a primary donor star with $M_{1}\!=\!19.95~M_{\odot}$ and a companion with a mass ratio of $q=0.6$ ($M_{2}\!=\!11.97~M_{\odot}$), in a circular orbit with an initial period of $P\!=\!3.98$~days. Both stars were initialized with a rotational velocity of $v_{\rm rot}=100$~\kms{}. We present the evolution of several key properties of this system in Fig.~\ref{fig_modelsMPA}.

Rotational effects are particularly important for the gainer. As shown by \citet{Packet81}, a star accreting material from a Keplerian disk can reach critical rotation after increasing its mass by only a few percent. The models of \citet{Jin+26} assume that accretion is inhibited once the star reaches critical rotation, following a so-called ``rotationally-limited accretion scheme''. These models include tidal interactions. However, they neglect the possible presence of accretion disks that could remove angular momentum and thereby allow accretion to continue \citep{Paczynski91,Popham&Narayan91}. Recently, \citet{Wang+21,Lechien+25,Sen+26} argued that this prescription may impose an overly restrictive limit on the amount of mass that can be accreted. The gainer may therefore accrete more mass than predicted by these models (middle panel of Fig.~\ref{fig_modelsMPA}).

For the analyzed system, mass transfer begins while the donor is still burning hydrogen in its core \citep[ Case A mass transfer as defined by][]{Kippenhahn&Weigert67}. As the donor is stripped, its surface becomes strongly He-enriched. Meanwhile, accretion spins up the gainer, limiting its accretion efficiency\footnote{At the same time, the resulting increase in rotational velocity triggers rotational mixing that affects the surface abundances \citep[e.g.][]{Renzo&Gotberg21}.} (middle panel of Fig.~\ref{fig_modelsMPA}). After the donor exhausts the hydrogen in its core, it detaches from its Roche lobe and mass transfer temporarily ceases. This first mass-transfer episode reverses the mass ratio, making the initially less massive star the more massive component of the system. 

Later, the donor develops hydrogen-shell burning, which drives its expansion and triggers a second mass-transfer episode, usually classified as Case~AB \citep[e.g. Fig.~3 in][]{Marchant&Bodensteiner24}. At this stage, the transferred material comes from layers that have already undergone intense nuclear processing and is strongly enriched in helium and heavier elements. The donor becomes a stripped star and rapidly evolves bluewards, undergoing core helium burning at very high effective temperatures (\Teff{},$\sim!100$~kK). Shortly after exhausting its helium core --a few thousand years-- it proceeds through the remaining nuclear-burning stages before ultimately undergoing a SN explosion. In the models of \citet{Jin+26} that do not encounter numerical issues or unstable mass transfer, the binary is assumed to be disrupted at the end of the primary’s evolution (except in cases where the primary becomes an AGB star or a white dwarf; see App.~B of \citealt{Jin+26} for details).
However, previous to that, its orbital period reaches a value within the range observed in our sample. Systems with lower initial mass ratios evolve towards even shorter orbital periods. In the limit, very low mass ratios ($q\lesssim0.4$) may lead to unstable mass transfer, resulting in a common-envelope phase or even a merger \citep[see][]{Marchant&Bodensteiner24}.

After the final detachment, thermohaline mixing becomes more efficient than rotational mixing in the accretor's outer layers \citep{Renzo&Gotberg21}. This significantly reduces the surface helium abundance of the accretor, which had reached values similar to those of the donor during the accretion phase (upper panel in Fig.~\ref{fig_modelsMPA}). Nevertheless, the final surface composition remains He-enriched compared to the initial He-abundance (dotted and dashed-dotted lines in the upper panel, respectively), in qualitative agreement with the observations. A higher accretion efficiency \citep[e.g.][]{Lechien+25,Xing+26} would presumably result in stronger He enrichment.

\section{Projected rotational velocity distribution}\label{App_vsini}

In a binary system, there might be two major processes affecting the evolution of the rotation of the stellar components beyond the stellar evolution: tidal interaction and mass accretion. The former tends to synchronize the rotation of
the stars with the orbit in systems; the latter exchange angular momentum from the donor to the accretor \citep{deMink+13}.

\begin{figure*}[!t]
\includegraphics[width=1.\hsize,trim={0 1 0 0}]{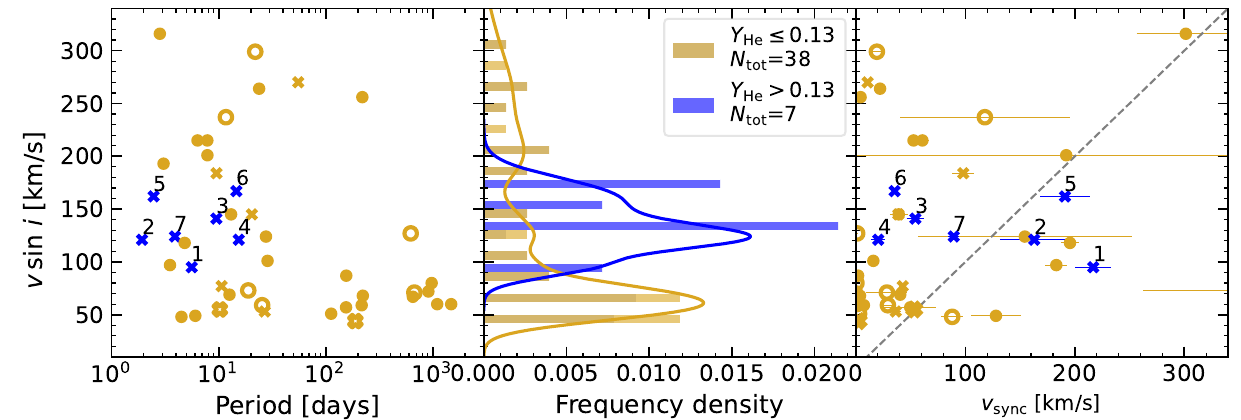}
  \caption{Left panel: Projected rotational velocity distribution as a function of the period for He-rich and He-normal systems (blue and yellow distributions, respectively). 
  Middle panel: Distribution of projected rotational velocity; lighter yellow bins for SB1d systems. Curves correspond to the empirical probability distribution computed as the addition of the probability distribution of the independent systems.
  Right panel: Comparison between the observed projected rotational velocity and the synchronous equatorial velocity. The stellar radii used to compute $v_{\rm sync}$ were obtained following \citet{Holgado+25}. For reference, the one-to-one relation is shown as a dashed line.}\label{fig_vsini}
\end{figure*}

In fig.~\ref{fig_vsini}, we present the \vsini{} distribution as a function of the period (left), its frequency density\footnote{i.e. the area under the distribution integrates to 1} (middle) separated according to their He-abundance in bins of 20~\kms{}, and its comparison with the expected synchronous equatorial velocity ($v_{\rm sync}$; right). In the former, same coding as in Fig.~\ref{fig_sHR}. In the histogram, bins in light yellow corresponds to SB1d systems (App.~\ref{app_disentangling}). Together with the histogram, we present the empirical probability distribution as the addition of the probability distribution of the independent systems (this has the advantage of being unsensitive both to the bin size and their edges).

In the left panel we note a different \vsini{} distribution depending on the period. Systems with $P<40$~days reach projected rotational velocities of $\sim250$~\kms{} (with three systems with even higher velocities), while systems with longer periods are concentrated at \vsini{}$\lesssim100$~\kms{}, with only three cases with higher velocities (1 with \vsini{}$\sim130$~\kms{} and 2 with \vsini{}>250~\kms{}). 

In the middle panel we find an asymmetric distribution for He-normal stars (yellow line), with a main peak at \vsini{}\,$\sim\!60$~\kms{} \citep[consistent with the initial velocity distribution in Galactic O-type stars reported in ][]{Holgado+22} and a tail of faster rotators reaching velocities above $\sim300$~\kms{}. 
In contrast, for He-rich stars (blue line) we find a more symmetric distribution centered at \vsini{}$\sim130$~\kms{} . The main peak of this distribution coincides with a secondary peak in the distribution of He-normal stars. While we note the possible common origin of both peaks, the low statistics prevents us to further support this hypothesis.

\subsection{The lack of fast rotators He-rich SB1 systems}

He-rich stars cluster around short period binaries with \vsini{}\,$\sim130$~\kms{}. While this value is  considerably faster than the birth velocity \citep[][]{Holgado+22}, it is still far from critical or fast rotation velocities (typically \vsini{}$\,\gtrsim\!200$~\kms{}). In this context, binarity provides a natural explanation, as tidal forces would produce a rotational synchronization in systems that are close enough \citep[][and references therein]{deMink+13,Langer+20}, either preventing them from reaching critical rotation during mass accretion events or accelerating them \citep[see, however, ][]{Lechien+25,Sen+26}.

We note that four of the He-rich systems have \vsini{} values slightly exceeding the expected $v_{\rm sync}$ (right panel of Fig.~\ref{fig_vsini}). Three of these systems have relatively long orbital periods ($P\gtrsim5$~days), for which tidal synchronization is expected to be less efficient. Nevertheless, they remain far from critical rotation, suggesting the presence of an efficient mechanism that prevents them from reaching critical rotational velocities \citep[see the discussion and models in][]{Renzo&Gotberg21}.
For the systems with \vsini{} lower than $v_{\rm sync}$, projection effects may conceal their true equatorial rotational velocities. We emphasize that all of these systems have short orbital periods ($P\lesssim5$~days), where tidal forces are expected to drive efficient synchronization \citep[][]{Zahn75}.
Interestingly, a similar characteristic velocity is also associated with the secondary peak in the \vsini{} distribution of He-normal stars (middle panel of Fig.~\ref{fig_vsini}), suggesting that these objects may share a common physical origin.

In wide systems, tidal synchronization is not efficient and can hardly prevent the spin up of the gainers during mass accretion or synchronize it with the companion \citep[e.g.][]{deMink+13}. However, we do not find any He-rich fast rotator system in our sample. Moreover, \cite{Simon-Diaz+26} reported a low number of them among the total IACOB sample. This could be related with (i) a harder detection of binaries with fast rotators and longer periods (Sect.~\ref{sec_longPeriod}) or (ii) a real lack of these systems. The latter could be the result of the disruption of these binaries \citep[e.g.][]{deMink&Langer&Izzard11}. This would naturally explain the larger fraction of RW reported in He-rich fast rotators. In addition, \cite{Carretero-Castrillo+26} found a lack of fast rotator runaways --produced in the BSS-- in binaries. Therefore, their work would reinforce the scenario of the disruption of these systems after the SN. 

Alternatively, the lack of He-rich fast rotators could result from the spin-down of the accretor following the mass-transfer episode. Such spin-down may be driven by tidal interactions, stellar winds, or the inward transport of angular momentum. In this context, some binary evolution models predict that rapidly rotating accretors experience enhanced angular-momentum loss through their stellar winds \citep{deMink+13}. \citet{Gagnier+19} investigated the interplay between wind-driven mass and angular-momentum losses and stellar rotation. They found that angular-momentum loss increases with the ratio of the stellar angular velocity to the critical angular velocity, even in cases where the global mass-loss rate may decrease in the single-wind regime. However, they cautioned that some of their assumptions --namely the absence of mass flux through the photosphere, a zero normal velocity imposed at the stellar surface, and vertical hydrostatic equilibrium in the surface layers-- may be too simplistic to accurately describe the surface layers of a wind-emitting massive star. These assumptions could affect the surface opacity and, consequently, the conditions for critical rotation, particularly in the most massive stars. Moreover, \cite{Renzo&Gotberg21} argued that the wind mass-loss rates may be overestimated, based on models reproducing the evolution of the accretor in $\zeta$~Ophiuchi. They found that the accretor reaches nearly rigid rotation by the end of the mass-transfer phase, which persists until the surface is spun down by stellar winds and evolutionary expansion. Nonetheless, they predict that the accretor retains a rotational velocity of $\sim100$~\kms{} by the end of the main sequence, which might be higher for lower mass losses.

A generalized spin-down scenario is difficult to reconcile with the higher incidence of fast rotators among apparently single He-rich O-type stars reported by \cite{Simon-Diaz+26} (previous paragraph). Future studies of these stars, together with the possible identification of the currently missing long-period He-rich systems, will help discriminate between these scenarios.

\end{appendix}

\begin{acknowledgements}
We acknowledge the anonymous referee for the careful reading of the manuscript and for the valuable suggestions that have helped improve its quality.
This work is part of grant CEX2025-001609-S, awarded to the Instituto de Astrofísica de Canarias under the Severo Ochoa Centre of Excellence program and funded by MICIU/AEI/10.13039/501100011033.
C.~M.-S., G.~H, and S.~S.-D. acknowledges the support from the Agencia Estatal de Investigación (AEI) Spanish Ministerio de Ciencia, Innovación y Universidades (MICIU) and the Fondo Europeo de Desarrollo Regional (FEDER) under grant Productos de la interacción de estrellas masivas reverlados por grandes sondeos espectroscópicos,  with references PID2024-159329NB-C21. The project leading to this application has received funding from Comisión Europea (EC) under Project OCEANS - Overcoming challenges in the evolution and nature of massive stars, HORIZON-MSCA-2023-SE-01, No G.A 101183150 Funded by the Unión Europea. 
R.~G. acknowledges support from grant PICT 2019-0344 and UNLP G189.
O.~G.~B. is a member of the Carrera del Investigador Científico of the Comisión de Investigaciones Científicas of the Provincia de Buenos Aires (CICPBA), Argentina.
E.~A.~S acknowledges support by the Spanish Agencia estatal de investigaci\'on via PID2021-124879NB-I00 and PID2024-161863NB-I00.
G.~H. received the support from the “La Caixa” Foundation (ID 100010434) under the fellowship code LCF/BQ/PI23/11970035.
J.~M.~A. acknowledges support from the Spanish Government Ministerio de Ciencia e Innovación and Agencia Estatal de Investigación (10.13039/501100011033) through grant PID2022-136640~NB-C22.
The IACOB survey made use of the Nordic Optical Telescope, and the Mercator Telescope at the Observatorio del Roque de los Muchachos (La Palma, Spain), and 2.2 m telescopes at Observatorio de La Silla (Chile).
C.~M.-S. acknowledges Y. Götberg, J. I. Arias, and the stellar group at the UNLP and IALP (Argentina) for the constructive discussions related to the topic of this Letter, as well as \textit{la Cande} for her valuable suggestions regarding the formatting of the manuscript.
O.~G.~B also thanks Juan Ignacio Rodriguez for his valuable help in handling the grid of models employed in the analysis presented in Appendix~C. 
\vspace{-0.5cm}
\end{acknowledgements}

\end{document}